\documentclass[letterpaper,twocolumn,preprintnumbers,amsmath,amssymb,showpacs]{revtex4}

\usepackage{xspace}
\usepackage{latexsym}
\usepackage{hyperref}

\newcommand{\ie}{i.e.,\xspace}

\newcommand{\ket}[1]{|{#1}\rangle} 
\newcommand{\lhv}{\ensuremath{\text{lhv}}}
\newcommand{\cov}{\ensuremath{\operatorname{cov}}}
\renewcommand{\vec}{}

\newtheorem{theorem}{Theorem}
\newtheorem{definition}{Definition}

\newenvironment{proof}[1]{%
  \begin{trivlist}{}{\setlength{\topsep}{0cm}\setlength{\partopsep}{0cm}}
  \item \textbf{#1.\@}\hspace*{1ex}\ignorespaces}%
  {\makebox[0cm]{}\nolinebreak\hfill$\Box$\end{trivlist}}

\begin{document}

\title{Combinatorics and Quantum Nonlocality}
\author{Harry Buhrman}
\thanks{Supported in part by the EU fifth framework projects QAIP,
IST-1999-11234, and RESQ, IST-2001-37559.}
\affiliation{CWI and University of Amsterdam,
P.O. Box 94079,
1090 GB Amsterdam, The Netherlands
\addtocounter{footnote}{1}
}
\author{Peter H{\o}yer}
\thanks{Supported in part by the Alberta Ingenuity Fund and
the Pacific Institute for the Mathe\-ma\-tical Sciences.}
\affiliation{Dept.{} of Comp.~Sci., University of Calgary,
             2500 University Drive N.W., Calgary AB, Canada T2N 1N4}
\author{Serge Massar}
\altaffiliation[Also at]{
Ecole Polytechnique, C.P. 165, 
Universit\'e Libre de Bruxelles, 1050 
Brussels, Belgium}
\thanks{supported in part by the EU fifth framework projects EQUIP, IST-1999-11053, and RESQ, IST-2001-37559.}
\affiliation{Service de Physique Th\'eorique,
Universit\'e Libre de Bruxelles,
 C.P. 225, Bvd. du Triomphe, 1050
Bruxelles, Belgium} 
\author{Hein R\"ohrig}
\thanks{Supported in part by the EU fifth framework projects QAIP,
IST-1999-11234, and RESQ, IST-2001-37559.}
\affiliation{CWI,
P.O. Box 94079,
1090 GB Amsterdam, The Netherlands
}

\date{February 4, 2003}

\begin{abstract}
  We use techniques for lower bounds on communication to derive
  necessary conditions (in terms of detector efficiency or amount of
  super-luminal communication) for being able to reproduce the quantum
  correlations occurring in EPR-type experiments with classical local
  hidden-variable theories. As an application, we consider $n$ parties
  sharing a GHZ-type state and show that the amount of super-luminal
  classical communication required to reproduce the correlations
  is at least $n(\log_2 n -3)$ bits and the maximum detector
  efficiency $\eta_*$ for which the resulting correlations can still
  be reproduced by a local hidden-variable theory 
  is upper bounded by $\eta_* \leq 8/n$ and thus decreases with~$n$.
\end{abstract}


\maketitle

\section{Introduction}

Some 40 years ago Bell showed that entangled quantum states are
nonlocal \cite{bell:epr}.  Although much is known about nonlocality in
the simplest cases, there are few clear results when the number $n$ of
systems or the dimension $d$ of the systems is large.  In this paper
we show how combinatorial techniques can be used to study quantum
nonlocality, particularly in the asymptotic cases.

We consider two ways in which the nonlocality of the quantum
correlation can be measured: 1)~the minimum efficiency $\eta$ of
detectors required for the correlations to exhibit nonlocality (this
is the so-called ``detection loophole''\cite{pearle}); 2)~the amount
of classical communication required to reproduce the quantum
correlations \cite{bct:simulating,steiner:quantifying}.  (Note that if
the measurements that are carried out on the entangled state are
spatially separated events, this communication would have to occur
faster than the speed of light, which is impossible.  If the events
are timelike separated, this communication can occur without
contradicting relativity.)  It has been shown that these two measures
are closely related \cite{gisingisin, massar:closing}. In this paper
we further explore this connection.

The amount of (possibly super-luminal) classical communication
required to reproduce the quantum correlations, is related to the
question examined in \cite{cb:substituting,bdht:multiparty} of whether
entanglement could decrease the amount of communication required to
compute a distributed function. This is the subject of the field of
computer science called ``communication complexity.'' Special
combinatorial techniques have been developed to prove lower bounds on
the amount of communication required to solve such problems, see
\cite{kushilevitz&nisan:cc} for an overview.  Here we show that the
same combinatorial techniques can be used to provide bounds on the
minimum detector efficiency required to close the detection loophole.

As an application of our method we obtain new results concerning how
nonlocality scales with the number of systems. We consider a specific
case in which $n$ parties each possess a two-dimensional system.  The
state of the overall system is a GHZ-type state. We show that the
amount of communication required to reproduce these correlations
classically is at least $n(\log_2 n -3)$ bits and that the minimum
detector efficiency $\eta_*$ required to close the detection loophole
in this case {\em decreases} as $1/n$ where $n$ is the number of
parties. This bound on $\eta_*$ is a significant improvement over
previous results, which could only show that the threshold detector
efficiency stays constant as the number of parties increases
\cite{mermin:extreme,zukowski,larsson01:_stric_claus_horne}. We
discuss possible experimental tests of these correlations.

\section{Definitions}

Consider the following situation.  There are $n$ spatially separated
parties; party $i$ receives an input $X_i \in \{ 1, \ldots, k \}$ and
produces an output $a_i \in \{ 1, \ldots, d \}$.  With $\vec X = (
X_1, \ldots, X_n )$ and $\vec a = (a_1, \ldots, a_n )$, let $P(\vec a
| \vec X)$ denote the probability of output $\vec a$ given input $\vec
X$.  We formalize this situation as follows.
\begin{definition}
  An \emph{$(n, k, d)$ correlation problem with pro\-mise $D \subseteq
    \{ 1, \ldots, k \}^n$} is a family of probability distributions
  $P( \cdot | \vec X)$ on the ``outputs'' $\{ 1, \ldots, d \}^n$, for
  each ``input'' $\vec X \in D$.
\end{definition}
In an experimental setup, we would configure the random-number
generators for the inputs to produce settings only from $D$ and thus
an adversary constructing a classical model would be ``promised'' that
he only has to reproduce the conditional probabilities for inputs from
$D$. A restriction of this type often simplifies the analysis of what
quantum and classical models can or cannot achieve.

Of particular interest are correlation problems obtained from quantum
mechanical experiments in which the parties share an entangled state
$\ket \psi$; each input $X_i$ determines a von~Neumann measurement
$\hat X_i$.  The measurement of $\hat X_i$ produces result $a_i$. The
probability $P_{\text{QM}}(\vec a | \vec X)$ to obtain outcome $\vec
a$ given input $\vec X$ is easily computed using the standard rules of
quantum mechanics.

In the case of inefficient detectors we enlarge the space of outputs
to $a_i \in \{ 1, \ldots, d \}\cup \{ \perp \}$, where $a_i = \perp$
is the event that the $i$\/th detector does not produce an output
(``click''). We suppose that each measurement $\hat X_i$ has
probability $\eta$ of giving a result and a probability $1-\eta$ of
not giving a result. We are interested in the case where all $n$
detectors give a result. This occurs with probability~$\eta^n$.

Let us consider classical models in which the parties cannot
communicate after they have received the inputs (such models are
called {\em local}).  The best the parties can do in this case is to
randomly select in advance a deterministic strategy. This motivates
the following definition.
\begin{definition}\label{def:detlhv}
  A \emph{deterministic local hidden variable (\lhv{}) model} is a
  family of functions $\lambda= (\lambda_1,\ldots,\lambda_n)$ from the
  inputs to the outputs: $\lambda_i : \{ 1, \ldots, k \} \rightarrow
  \{ 1, \ldots, d, \perp \}$.  Each party outputs $a_i =
  \lambda_i(X_i)$.
  
  A \emph{probabilistic \lhv{} model} (or just \emph{\lhv{} model}) is
  a probability distribution $\nu(\lambda)$ over all deterministic
  \lhv{} models for given $(n, k, d)$.
\end{definition}
Thus in probabilistic \lhv{} models the parties first randomly choose
a deterministic \lhv{} model $\lambda$ using the probability
distribution $\nu$. Each party then outputs $a_i = \lambda_i(X_i)$.
In general, \lhv{} models
cannot reproduce the quantum correlations $P_{\text{QM}}$ except if
the detector efficiency $\eta$ is sufficiently small. For a fixed
correlation problem (based on the quantum correlations
$P_{\text{QM}}$), we denote by $\eta_* = \eta_* (P_{\text{QM}})$
the maximum detector efficiency for which a \lhv{} model exists.  Our
aim is to obtain upper bounds on $\eta_*$.

We also consider classical models with communication. In such
models, the parties may communicate over a (possibly super-luminal)
classical broadcast channel in order to reproduce the quantum
correlations $P_{\text{QM}}$.  Different communication models exist
depending on whether the parties do not have access to randomness,
possess local randomness only, or possess shared randomness.
\begin{definition}
  For a correlation problem $P$, denote by $R(P)$, and
  $R^{\text{pub}}(P)$, respectively, the minimum number of bits that
  must be broadcast in order to perfectly reproduce the correlations
  $P$ when the parties are have local randomness only
  or have shared randomness, respectively.
\end{definition}
Clearly, $R(P) \geq R^{\text{pub}} (P)$. $R(P)$ can be
infinite \cite{MBCC}; in this work we derive lower bounds on
$R^{\text{pub}}$.

Since the results of quantum measurements are inherently random, it is
in general impossible to reproduce the quantum correlations using
deterministic \lhv{} models or using deterministic models with
communication.  However, deterministic models are a very useful tool
for studying the probabilistic models. In particular, properties of
\emph{all} deterministic models necessarily also hold for \emph{all}
probabilistic models.

\section{Combinatorial Bounds}

Our results are not based on the details of the probability
distribution $P(\cdot|X)$, but only on whether a given output $a$ has
nonzero probability given an input $X$.
\begin{definition}
  For a given correlation problem with promise $D$, we
  call output $a$ \emph{admissible} on input $X \in D$ if and only if
  $P(a | X) > 0$.
  
  We call a \lhv{} model \emph{error free} if and only if it produces
  for all inputs only admissible outputs.
\end{definition}
In the following we consider only error-free models.
We now introduce some definitions and notation which allow us to state
our main result.

For sets $A_1$, \ldots, $A_n$, a subset $R$ of the Cartesian product
$A_1 \times \cdots \times A_n$ is called a \emph{rectangle} if there
are $R_1 \subseteq A_1$, \ldots, $R_n \subseteq A_n$ such that $R =
R_1 \times \cdots \times R_n$, \ie $R$ is a Cartesian product itself.

The importance of rectangles is that
for a deterministic \lhv{} model $\vec \lambda = (\lambda_1, \ldots,
\lambda_n)$, the set $ R_{\vec \lambda} (\vec a) = \{ \vec X
\mathrel{:}\vec \lambda (\vec X) = \vec a\} $ of all inputs $\vec X$
leading to output $\vec a$ is a rectangle: $R_{\vec \lambda}( \vec a)
= \lambda_1^{-1} ( a_1) \times \cdots \times \lambda_n^{-1} ( a_n)$.

Let us consider outputs $\vec a \in
\{1,\ldots,d\}^n$. We call a set 
$R \subseteq \{1, \ldots, k \}^n$ \emph{$\vec
  a$-monochromatic} with respect to an $(n,k,d)$ correlation problem
$P$ with promise $D$ if for all $\vec X \in R \cap D$, $P(\vec a|\vec
X) > 0$ (the term ``monochromatic'' derives from interpreting $a$ as a
color label of the points in $X$).  Monochromatic rectangles play a
central role in what follows because
a deterministic \lhv{} model $\vec
\lambda$ produces only admissible outputs if and only if for all $\vec
a\in
\{1,\ldots,d\}^n$, $R_{\vec \lambda} (\vec a)$ is $\vec
a$-monochromatic.  

Our results are stated in terms of maximum size of monochromatic
rectangles. Specifically, for a correlation problem $P$
with promise $D$, we will denote the size of the largest overlap of
the promise $D$ with an $a$-monochromatic rectangle (with $\vec a \in
\{1,\ldots,d\}^n$) by
\[
r(P) := \max \{ | R \cap D | \mathrel{:}  R \text{ monochromatic rectangle} \}
.
\]
This allows us to bound the minimum number $\cov(P)$ of monochromatic
rectangles required to cover $D$: $ \cov(P) \geq |D|/r(P)$. Our main
results can now be stated with precision.

\begin{theorem}\label{thm:effrect}
  The maximum detection efficiency $\eta_*$ for which an $(n,k,d)$
  correlation problem $P$ with promise $D$ can be explained by a
  \lhv{} theory is bounded by
  \begin{equation*}
    \eta_* \leq d \left( \frac{ r(P) }{ |D| } \right)^{1/n} 
    .
  \end{equation*}
\end{theorem}

\begin{theorem}\label{thm:sharedRandCommRect}
  In the communication model where the parties have shared randomness,
  the minimum amount of communication required to reproduce the
  correlations for an $(n,k,d)$ correlation problem $P$ with promise
  $D$ 
is 
  \[
  R^{\text{pub}}(P) \geq \log_2 \left( \frac{ \cov(P) }{ d^n}  \right)
  \ge \log_2 \left( \frac{ |D| }{ d^n r(P) } \right) 
  .
  \]
\end{theorem}
Thus the same quantities provide a bound for the minimum amount of
classical communication required to reproduce the quantum correlations
and the minimum detector efficiency $\eta_*$ required for the
correlations to be nonlocal.

\begin{proof}{Proof of Theorem~\ref{thm:effrect}}
  A general \lhv{} model without error can be interpreted as a
  probability distribution $\nu ( \lambda)$ over all deterministic
  error-free \lhv{} models $\lambda$: for each run, a deterministic
  \lhv{} model $\lambda$ is chosen according to the distribution $\nu$
  before the inputs are received. Let
  $A_{\lambda,X} = 1$ if $\lambda_i (X_i) \ne \perp$ for all
  $i=1,\ldots,n$, and $A_{\lambda,X}=0$ otherwise; \ie
  $A_{\lambda,X}=1$ if and only if all $n$ detectors ``click'' in the
  deterministic \lhv{} model $\lambda$ on input $X$. With this
  notation, the probability that the probabilistic \lhv{} model
  produces an output for inputs drawn uniformly at random from $D$ is
  $q := \sum_X 1/|D| \sum_\lambda \nu(\lambda) A_{\lambda,X}$.  If the
  detector efficiency is $\eta$, then the probability of all detectors
  clicking is $\eta^n = q$.

  If we can show that any deterministic
  \lhv{} model produces an admissible output for at most $d^n r(P)$
  inputs, then no probabilistic \lhv{} model can produce an admissible
  output with probability $q > d^n r(P) / |D|$, because $q =
  \sum_\lambda \nu(\lambda) \sum_X A_{\lambda, X}/|D| \le
  \sum_{\lambda} \nu(\lambda) d^n r(P) / |D| = d^n r(P) / |D|$.
  
  A deterministic \lhv{} model $\vec \lambda$ that produces only
  admissible outputs can give an output with all detectors clicking
  for at most $d^n r(P)$ input settings, because for each of $d^n$
  possible outputs $a$, $R_\lambda (a)$ is a monochromatic rectangle
  of size $|R_\lambda(a)| \le r(P)$.

  In conclusion, if $d^n r(P)/|D| < q = \eta^n$
  no \lhv{} model can reproduce the correlations.
\end{proof}

\begin{proof}{Proof of Theorem~\ref{thm:sharedRandCommRect}}
  We now consider classical models with a limited amount of
  communication and perfect detector efficiency.  As in the previous
  proof, any such model can be interpreted as a probability
  distribution $\nu (\lambda)$ over deterministic models $\lambda$
  with communication and perfect detector efficiency.  Since there are
  no errors, the models $\lambda$ with nonzero probability
  $\nu(\lambda) >0$ in this decomposition must produce only admissible
  outputs. We now show that deterministic models that produce only
  admissible outputs require at least $\log_2 ( \cov(P) / d^n )$ bits
  of communication. Hence probabilistic models also require at least
  $\log_2 ( \cov(P) / d^n )$ bits of communication.
  
  Consider a deterministic model with at most $c$ bits of
  communication producing only admissible outputs. This model can be
  represented by a tree where each node is labeled by the party whose
  turn it is to send a message, and the edges to children are labeled
  by which message was sent; a run of the model corresponds to a path
  from the root of the tree to a leaf. The communication tree has at
  most $2^c$ leaves; with each leaf $l$ we associate $d^n$ sets $R(l,
  \vec a) \subseteq \{ 1, \ldots, k \}^n$ of the inputs leading to
  this leaf and producing output $\vec a$.  The deterministic model
  produces only admissible outputs, hence, $R(l, \vec a)$ is a
  monochromatic rectangle for each $l$ and $\vec a$. Furthermore, the
  $R(l, \vec a)$ must cover $D$ since every input produces an output
  (the detector efficiency is 1), therefore we have $2^c d^n \ge
  \cov(P)$ or, equivalently, $c \ge \log_2 ( \cov(P) / d^n )$.
\end{proof}
The proofs of Theorems~\ref{thm:effrect} 
and~\ref{thm:sharedRandCommRect} show the close relation between
the detection loophole and the amount of communication required in the
shared randomness model. In both cases the size of the largest
monochromatic rectangle yields bounds on the detector efficiency and
the amount of communication, respectively.

We can map any communication model with $c$ bits of communication with
shared randomness into a model with inefficient detectors with
efficiency $ \eta^n = 2^{ -c } $: the shared randomness is used to
randomly select a conversation between the parties and each party $i$
checks whether its input $X_i$ is compatible with the conversation
and, if yes, produces output $a_i$ according to the 
communication model and
otherwise produces no output. The probability that all detectors click
is equal to the probability that $\vec X$ belongs to the conversation.
Since each input belongs to one and only one conversation, the
probability that all detectors click is equal to one over the number
of conversations.  Note that in this model the probability that a
specific detector, say detector $i$, clicks may depend on the input
$X_i$ (however, the probability that all detectors click remains
independent of the input).

\begin{theorem}\label{thm:outputForComm}
  Consider \lhv{} models where the probability that all detectors
  click is independent of the input, but where the probability that
  each detector clicks, say detector $i$, may depend on its input
  $X_i$.  Then there exists a \lhv{} model if the probability $\eta^n$
  that all detectors click is at most $2^{-R^{\text{pub}}}$. This
  implies that in these models,
  \begin{equation}
    \eta_*^n \geq 2^{-R^{\text{pub}}}\ .
    \label{model}
  \end{equation}
\end{theorem}
Theorem~\ref{thm:effrect} also applies to the models considered in
Theorem~\ref{thm:outputForComm}. Observe that together with
Theorem~\ref{thm:sharedRandCommRect}, this implies that in models
where the bounds on $R^{\text{pub}}$ coming from rectangles are almost
tight, eq.~(\ref{model}) also is almost tight. This is for instance
the case in the model considered below (see the discussion in
\cite{bdht:multiparty}).

\section{An Application}

We now apply Theorems~\ref{thm:effrect} and
\ref{thm:sharedRandCommRect} to an example inspired by
\cite{bdht:multiparty}.
\begin{theorem}\label{thm:multiparty}
  For each $n \ge 3$, there is a correlation problem $P$ 
  with $n$ parties,
  $k = 2^{\lceil \log_2 n \rceil} \in 
  \Theta ( n )$ possible inputs per party, and $d = 2$ possible
  outputs per party with the following properties,
  \begin{itemize}
  \item it can be perfectly reproduced by a quantum mechanical system
    sharing a GHZ-type state $\ket \psi = ( \ket{ 0^n } +
    \ket{ 1^n }) / \sqrt{2}$;
  \item it cannot be reproduced by an error-free \lhv{} model for
    detector efficiency $\eta > 8/n$ or with 
    less than $n ( \log_2 n -
    \textup{constant})$ bits of communication.
  \end{itemize}
\end{theorem}
\begin{proof}{Proof}
  Consider $n$ parties, each of which possesses a two-dimensional
  quantum system. The entangled state of the $n$ systems is a GHZ-type
  state,
  $
  \ket \psi = (\ket{0^n} + \ket{1^n} )/\sqrt{2}
  $.
  Each party $i$ receives as input a $l$ bit string: $X_i
  \in\{0,\ldots,2^l-1\}$. The measurements carried out by the parties
  are the following. Each party carries out the unitary transformation
  $\ket 0 \mapsto \ket 0$, $\ket 1 \mapsto \exp(2\pi\sqrt{-1} X_i / 2^l)
  \ket 1$. Each party then measures an operator whose eigenstates are
  $(\ket 0 + \ket 1)/\sqrt{2}$ and $(\ket 0 - \ket 1)/\sqrt{2}$. The
  first outcome is assigned the value $a_i=0$, the second
  the value $a_i=1$, so $d=2$.

  In this measurement scenario, we have that if
  \begin{equation}\label{PP}
    \left ( \sum_{i=1}^n X_i \right) \bmod 2^{l-1} = 0 , 
  \end{equation}
  then
  \begin{equation}\label{PC}
    \left ( \sum_{i=1}^n a_i \right) \bmod 2 = \frac{1}{2^{l-1}}
      \left \{ \left ( \sum_{i=1}^n X_i \right) \bmod 2^{l} \right \}. 
  \end{equation}
  The promise $D$ is taken to be the set of $\vec X$ that satisfy
  (\ref{PP}); the size of $D$ is $|D| = 2^{(n-1) l}$ (since in
  eq.~(\ref{PP}) $n-1$ of the $X_i$ can be freely chosen and the last
  is then fixed). We say $S \subseteq \{ 0,\ldots,2^l-1\}^n$ is
  $b$-monochromatic if 
  \[
  \frac{1}{2^{l-1}} 
  \left \{ \left ( \sum_{i=1}^n X_i \right) \bmod 2^{l} \right \} 
  = b
  \]
  for all $X \in S \cap D$.  Reference~\cite{bdht:multiparty} shows
  using addition theorems for finite groups that for $b \in \{ 0, 1
  \}$ any $b$-monochromatic rectangle $R$ has size bounded by
  \[
  r \leq \left( \frac{ 2^l- 2}{n} + 1 \right )^n 
  \]
  whenever $R \cap D \ne \emptyset$.
  By eq.~(\ref{PC}), for all $a \in \{ 0, 1 \}^n$, any
  $a$-monochromatic set is also $(\sum_i a_i \bmod 2)$-monochromatic.  
  Therefore $r(P) \le r$.  
  Substituting these values into the bound of
  Theorem~\ref{thm:effrect}, we obtain
  \begin{equation}\label{five}
    \eta_* \leq  {2 \cdot 2 ^{l/n}  }
    \left ( \frac{1}{n} - \frac{2}{n 2^l} + \frac{1}{2^l}\right)
    \leq  \frac{4}{n} n^{1/n} \leq \frac{8}{n},
  \end{equation}
  where $l$ is taken so that $\log_2 n \leq l < \log_2 n +1$ and 
  the last inequality is valid when $n\geq 3$.
  Similarly, Theorem~\ref{thm:sharedRandCommRect} yields
  $R^{\text{pub}} \geq n ( \log_2 n - 3)$.
\end{proof}

GHZ-type states involving three parties have been realized 
experimentally
in cavity QED \cite{cavity}, in optics using spontaneous parametric
down-conversion \cite{photon}, and in NMR \cite{nmr}.  
Optimizing eq.~(\ref{five}) 
shows that only a moderate number of parties is required
for the super-luminal communication cost $R^{pub}$ to increase and the
threshold detection efficiency to decrease significantly.  
For example, for $n=8$, $\eta_* \leq 0.46$ and $R^{pub}\geq 9$ bits, 
and for $n=12$,
$\eta_* \leq 0.29$ and $R^{pub}\geq 22$ bits.  Therefore it should be
feasible to realize experimentally the states and measurements
required for these tests of quantum nonlocality, using either
extensions of the techniques above, ion traps \cite{ion} (up to four
ions have been entangled), or atomic ensembles following a recent
proposal \cite{Duan}. However, in experiments noise is inevitable,
whereas we have considered here only noiseless correlations. In an
upcoming paper we will show how Theorem~\ref{thm:multiparty} can be
generalized in the presence of noise.

In summary, we show that the combinatorial techniques developed
to study communication complexity can be used to quantify the amount
of (super-luminal) communication required to reproduce the quantum
correlations or the minimum detection efficiency required to close the
detection loophole.  We apply this approach to multipartite
entanglement: building on a result of \cite{bdht:multiparty} we show
that in a precise and operational way the nonlocality increases with
the number of parties.

\vfill


\end{document}